\documentclass[review]{elsarticle}
\usepackage{lineno,hyperref}
\usepackage{graphicx}
\usepackage[linesnumbered,ruled,lined]{algorithm2e}
\usepackage{float}
\usepackage{tikz}
\usepackage{epstopdf}
\usepackage{amssymb}
\usepackage{amsmath}
\usepackage{subcaption}
\usetikzlibrary{arrows.meta}
\newcommand{\tr}{\mathop{\mathrm{tr}}}

\journal{J. X. X. X}

\begin{document}
	%
	%
	\begin{frontmatter}
		\title{A simple artificial damping method for 
			total Lagrangian smoothed particle hydrodynamics}
		\author[myfirstaddress]{Chi Zhang }
		\ead{c.zhang@tum.de}
		\author[myfirstaddress]{Yujie Zhu}
		\ead{yujie.zhu@tum.de}
		\author[mysecondaryaddress]{Yongchuan Yu}
		\ead{yongchuan.yu@tum.de}
		\author[myfirstaddress]{Massoud Rezavand}
		\ead{massoud.rezavand@tum.de}
		\author[myfirstaddress]{Xiangyu Hu \corref{mycorrespondingauthor}}
		\ead{xiangyu.hu@tum.de}
		\address[myfirstaddress]{Department of Mechanical Engineering, 
			Technical University of Munich, 85748 Garching, Germany}
		\address[mysecondaryaddress]{Department of Aerospace and Geodesy, 
			Technical University of Munich, 82024 Taufkirchen, Germany}
		\cortext[mycorrespondingauthor]{Corresponding author.}
		\begin{abstract}
			In this paper, 
			we present a simple artificial damping method to enhance the robustness of  
			total Lagrangian smoothed particle hydrodynamics (TL-SPH). 
			Specifically, 
			an artificial damping stress based on the Kelvin-Voigt type damper 
			with a scaling factor imitating a von Neumann-Richtmyer type artificial viscosity 
			is introduced in the constitutive equation to alleviate the spurious oscillation 
			in the vicinity of the sharp spatial gradients. 
			After validating the robustness and accuracy of the present method 
			with a set of benchmark tests with very challenging cases, 
			we demonstrate its potentials in the field of bio-mechanics by simulating the deformation of complex stent structures. 
		\end{abstract}
		
	\end{frontmatter}
	%
	%
	\section{Introduction}\label{sec:introduction}
	Numerical simulation of large-strain solid dynamics problems, 
	where flexible structures experience large deformations, 
	plays a key role in a vast range of engineering problems in the aerospace, 
	automotive, manufacturing and biomaterial industries. 
	Besides the traditional mesh-based finite element methods (FEM), 
	smooth particle hydrodynamics (SPH), 
	which is a mesh-free method and originally designed for fluid dynamics
	\cite{lucy1977numerical, gingold1977smoothed}, 
	has also been adopted for such problems \cite{libersky1991smooth,randles1996smoothed}, 
	and received increasing attention in the past decades. 
	Generally, there are two types of SPH formulation for solid dynamics:
	one is updated Lagrangian SPH (UL-SPH) formulation,  
	in which the current configuration is used as the reference
	\cite{swegle1995smoothed, johnson1996normalized, randles1996smoothed, 
		dilts1999moving, monaghan1994simulating, gray2001sph, dyka1997stress, zhang2017generalized}; the other is total Lagrangian SPH (TL-SPH) formulation \cite{vignjevic2006sph}, 
	in which the initial configuration is used as the reference. 
	Compared with UL-SPH, TL-SPH shows promising potential for solid dynamics 
	due to its attractive advantages in
	being free from tensile instability and ensuring $1$st-order consistency when computing deformation gradient by introducing a kernel correction matrix. 
	Since its inception, 
	TL-SPH has been applied for the problems of necking and fracture in electromagnetically driven rings \cite{de2013total}, 
	thermomechanical deformations \cite{ba2018thermomechanical}, 
	fluid-structure interaction (FSI) \cite{antoci2007numerical, khayyer2018enhanced, liu2019smoothed, zhang2020multi} 
	and bio-mechanics \cite{zhang2020sphinxsys, zhang2020integrative}, 
	among many others.  
	It is known that without appropriate stabilization 
	the original TL-SPH exhibits spurious fluctuations in the vicinity of sharp spatial gradients (as will also be shown in Section \ref{sec:cable}). 
	This deficiency may result in numerical instability and lead to wrongly predicted deformation for problems involving large strain. 
	To rectify this deficiency, Lee et al. \cite{lee2016new} proposed a Jameson-Schmidt-Turkel (JST) based method, 
	which shows good performance of eliminating spurious pressure oscillations 
	in nearly incompressible solids.
	In a recent work, Lee et al. \cite{lee2019total} further proposed 
	a new stabilization method by introducing 
	a characteristic-based Riemann solver in conjunction 
	with a linear reconstruction procedure. 
	This method also shows good performance in the simulation of nearly and truly incompressible explicit fast solid dynamics with large deformations. 
	
	The main objective of this paper is to present 
	a simple and effective artificial damping method for TL-SPH to enhance its numerical stability.
	In particular, 
	an artificial damping based on Kelvin–Voigt type damper is introduced 
	in the constitutive equation to alleviate the spurious oscillation 
	in the vicinity of the sharp spatial gradients. 
	By imitating a von Neumann-Richtmyer type artificial viscosity, 
	a scaling factor is introduced to control the damping force. 
	Compared with the works of Refs. \cite{lee2016new, lee2019total}, 
	the present method is very simple and easy to be implemented into 
	the original TL-SPH formulation in a straightforward way. 
	A set of benchmark tests with very challenging cases are investigated 
	to validate the robustness and accuracy of the present method.
	Furthermore, its versatility is demonstrated 
	by modeling the deformation of complex stent structures. 
	Also, all the codes and data-sets accompanying this work are available 
	from the open-source library of SPHinXsys \cite{zhang2020sphinxsys}
	on GitHub at \url{https://github.com/Xiangyu-Hu/SPHinXsys}. 
	The remainder of this paper is arranged as follows: 
	Section \ref{sec:kinematics} briefly summarizes 
	the governing equations and original TL-SPH formulation. 
	Then, TL-SPH-KV is detailed in Section \ref{sec:relaxation} and numerical validations and applications are presented and discussed in Section \ref{sec:example}.
	Finally, brief concluding remarks are given in Section \ref{sec:conclusion}.
	\section{Preliminary}\label{sec:kinematics}
	The kinematics of the finite-deformation solid dynamics can be characterized by introducing a deformation map $\varphi$, 
	which maps a material point $\mathbf{\mathbf{r}^0}$ from the initial reference configuration $\Omega^0 \subset \mathbb{R}^d $ 
	to the point $\mathbf{r} = \mathbf{\varphi}\left(\mathbf{r}^0, t\right)$ 
	in the deformed configuration $\Omega = \mathbf{\varphi} \left(\Omega^0\right)$. 
	Here, we denote the superscript $\left( {\bullet} \right)^0$ as the quantities in the initial reference configuration. 
	Then, the deformation tensor $\mathbb{F}$ can be defined as 
	\begin{equation} \label{eq:deformationtensor}
		\mathbb{F} = \nabla^{0} {\varphi} =  \frac{\partial \varphi}{\partial \mathbf{r}^0}  = \frac{\partial \mathbf{r}}{\partial \mathbf{r}^0} ,
	\end{equation}
	where the derivative is evaluated with respect to the initial reference configuration. 
	Note that the deformation tensor $\mathbb{F}$ can also be calculated from the 
	point displacement $\mathbf{u} = \mathbf{r} - \mathbf{r}^0$ through
	\begin{equation} \label{eq:deformationtensor-displacement}
		\mathbb{F} = \nabla^{0} {\mathbf{u}}  + \mathbb{I},
	\end{equation}
	where $\mathbb{I}$ represents the unit matrix. 
	\subsection{Governing equation for solid dynamics}\label{sec:governing-equation}
	In the total Lagrangian framework, 
	the momentum conservation equation can be expressed as
	\begin{equation}\label{eq:mechanical-mom}
		\rho^0 \frac{\text{d} \mathbf{v}}{\text{d} t}  =  \nabla^{0} \cdot \mathbb{P}^T  + \rho^0 \mathbf{g},
	\end{equation}
	where $\rho^0$ is the density in the reference configuration, 
	$\mathbf{v}$ the velocity and $\mathbb{P}$ the first Piola-Kirchhoff stress tensor. 
	For an ideal elastic or Kirchhoff material, $\mathbb{P}$ is given by
	\begin{equation}\label{linear-elasticity}
		\mathbb{P} = \mathbb{F} \mathbb{S}, 
	\end{equation}
	where $\mathbb{S}$ represents the second Piola-Kirchhoff stress which is evaluated via the constitutive equation relating $\mathbb{F}$ with 
	the Green-Lagrangian strain tensor $\mathbb{E}$ defined as 
	\begin{equation}\label{Lagrangian-strain}
		\mathbb{E} = \frac{1}{2} \left( \mathbb{F}^{T}\mathbb{F} - \mathbb{I}\right) .
	\end{equation}
	In particular, when the material is linear elastic and isotropic, the constitutive equation is simply given by
	\begin{eqnarray}\label{isotropic-linear-elasticity}
		\mathbb{S} & = & K \tr\left(\mathbb{E}\right)  \mathbb{I} + 2 G \left(\mathbb{E} - \frac{1}{3}\tr\left(\mathbb{E}\right)  \mathbb{I} \right) \nonumber \\
		& = & \lambda \tr\left(\mathbb{E}\right) \mathbb{I} + 2 \mu \mathbb{E} ,
	\end{eqnarray}
	where $\lambda$ and $\mu$ are Lam$\acute{e}$ parameters, 
	$K = \lambda + (2\mu/3)$ the bulk modulus and $G = \mu$ the shear modulus. 
	The relation between the two modulus is given by
	\begin{equation}\label{relation-modulus}
		E = 2G \left(1+2\nu\right) = 3K\left(1 - 2\nu\right),
	\end{equation}
	where $E$ denotes the Young's modulus and $\nu$ the Poisson's ratio. 
	Note that the sound speed of solid structure is defined as $c = \sqrt{K/\rho^{0}}$. 
	In the present work, 
	a neo-Hookean material model defined by the strain-energy density function
	\begin{eqnarray}\label{Neo-Hookean-energy}
		W  =  \mu \tr \left(\mathbb{E}\right) - \mu \ln J + \frac{\lambda}{2}(\ln J)^{2},
	\end{eqnarray}
	is also applied for predicting the nonlinear stress-strain behavior of materials undergoing large deformations. 
	For neo-Hookean material, the second Piola-Kirchhoff stress $\mathbb{S}$ can be derived as 
	\begin{equation}\label{2rd-PK}
		\mathbb{S} = \frac{\partial W}{\partial \mathbb{E}}.
	\end{equation}
	\subsection{TL-SPH formulation}\label{sec:tl-sph}
	In TL-SPH, the kernel correction or normalization technique \cite{vignjevic2000treatment, bonet2002simplified,randles1996smoothed} 
	has demonstrated its effects to improve the accuracy and consistency of SPH method. 
	The correction matrix $\mathbb B^0$ is introduced as \cite{vignjevic2006sph} 
	\begin{equation} \label{eq:sph-correctmatrix}
		\mathbb{B}^0_i = \left( -\sum_j V^0_j \mathbf r_{ij}^0 \otimes \nabla^0_i W_{ij} \right) ^{-1} ,
	\end{equation}
	where 
	\begin{equation}\label{strongkernel}
		\nabla^0_i W_{ij} = \frac{\partial W\left( |\mathbf{r}^0_{ij}|, h \right)}  {\partial |\mathbf{r}^0_{ij}|} \mathbf{e}^0_{ij},
	\end{equation}
	denoting the gradient of the kernel function evaluated at the initial reference configuration. 
	Note that the correction matrix is computed in the initial configuration and, therefore, 
	it is calculated only once before the simulation. 
	Then, the momentum conservation equation, Eq.\eqref {eq:mechanical-mom}, can be discretized as 
	\begin{equation}\label{eq:sph-mechanical-mom}
		\frac{\text{d}\mathbf{v}_i}{\text{d}t} = \frac{2}{m_i} \sum_j V^0_i V^0_j \tilde{\mathbb{P}}_{ij} \nabla^0_i W_{ij} + \mathbf{g},
	\end{equation} 
	where the inter-particle averaged first Piola-Kirchhoff stress $\tilde{\mathbb{P}}$ is defined as
	\begin{equation}
		\tilde{\mathbb{P}}_{ij} = \frac{1}{2} \left( \mathbb{P}_i \mathbb{B}^0_i + \mathbb{P}_j \mathbb{B}^0_j \right). 
	\end{equation}
	Here, 
	the first Piola-Kirchhoff stress tensor is computed with the constitutive law where the deformation tensor $\mathbb{F}$ is updated by the change rate evaluated through 
	\begin{equation}\label{rate-defmoration}
		\frac{\text d \mathbb F_i}{\text d t}  = \left( - \sum_j V^0_j \mathbf v_{ij}  \otimes \nabla^0_i W_{ij}  \right) \mathbb B^0_i .
	\end{equation}
	%
	\section{Kelvin-Voigt type damping}\label{sec:relaxation}
	An elastic solid undergoing large strains can be modeled with the mechanical components of springs and dashpots.
	The former represents the restorative force component and the later denotes the damping component. 
	The Kelvin-Voigt (KV) model can be represented by a purely viscous damper and purely elastic spring connected in parallel, and the total stress is decomposed into two parts 
	\begin{equation}\label{eq:kv-stress}
		\sigma_{total} = \sigma_{S} + \sigma_{D}. 
	\end{equation}
	Here, $\sigma_{total}$ is the total stress, 
	$\sigma_{S}$ the elastic stress 
	and $\sigma_{D}$ represent the damper stress as
	\begin{equation}\label{eq:kv-mdoel}
		\sigma_{D} =  \eta \frac{\text{d}\epsilon(t)}{\text{d}t}
	\end{equation}
	where $\text{d}\epsilon(t)/\text{d}t$ is the strain rate 
	and $\eta$ the physical viscosity. Applying the KV model to TL-SPH formulation, 
	the second Piola-Kirchhoff stress $\mathbb{S}$  can be rewritten as 
	\begin{equation}
		\mathbb{S} = \mathbb{S}_{S} + \mathbb{S}_D ,
	\end{equation}
	where $\mathbb{S}_{S}$ is given by the constitutive equation of Eq. \eqref{isotropic-linear-elasticity} or Eq. \eqref{Neo-Hookean-energy}, 
	and the damper $\mathbb{S}_D$ is defined as 
	\begin{equation}\label{eq:s-d}
		\mathbb{S}_D =  \pi \frac{\text{d}\epsilon(t)}{\text{d}t} 
		= \frac{\pi}{2}\left[\left( \frac{\text{d}\mathbb{F}}{\text{d}t} \right)^T \mathbb{F} + \mathbb{F}^T  \left( \frac{\text{d}\mathbb{F}}{\text{d}t}\right)  \right] , 
	\end{equation}
	where $\pi$ represents an artificial viscosity.
	As the main objective of introducing the KV-type damper is to enhance the robustness of original TL-SPH, 
	we introduce a von Neumann-Richtmyer type scaling factor with the speed of sound $c$ for Eq.  \eqref{eq:s-d} as 
	\begin{equation}\label{eq:artificial-v}
		\pi = \alpha \rho c h. 
	\end{equation}
	Here, $\alpha$ is a suitable and constant parameter and $h$ the smoothing length.
	Note that a similar parameter $\pi$ is also widely used in the artificial viscosity 
	for Eulerain shock-capturing schemes in modeling compressible flow.
	We suggest $\alpha = 0.5$ according to numerical experiments 
	and use it throughout this paper. 
	%
	%
	%
	\section{Numerical examples}\label{sec:example}
	In this section, 
	we first study three benchmark tests, 
	where the structures may experience large deformation, 
	to validate the robustness and performance of the proposed method (denoted as TL-SPH-KV).
	We also compare numerical results with those obtained by the original TL-SPH in which no damping stress is applied (denoted as “TL-SPH”). 
	Having the validation studies presented, 
	we then demonstrate the versatility of the method for applications in bio-mechanical system, 
	i.e. stent structures.
	In all the following examples,
	the $5th$-order Wendland smoothing kernel function with an smoothing length of $h= 1.15dp$ 
	is employed, 
	where $dp$ represents the initial inter-particle spacing.
	The position-based Verlet scheme, which is a two-step explicit algorithm proposed in the work of Zhang et al. \cite{zhang2020multi}, is used for time integration.
	The CFL number applied here is $ CFL = 0.6$, 
	which is twice as that used in the work of Lee et al . \cite{lee2019total}.
	\subsection{Wave propagation in a cable}\label{sec:cable}
	\begin{figure}[tb!]
		\centering
		\includegraphics[trim = 6cm 7cm 6cm 7cm, clip, width=\textwidth]{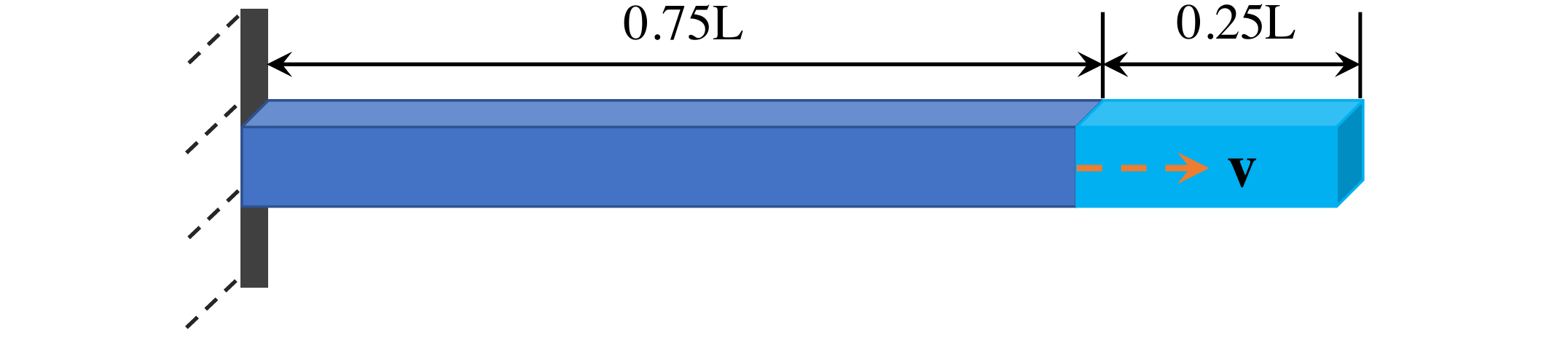}
		\caption{Wave propagation in a cable: Initial configuration.}
		\label{figs:cable-setup}
	\end{figure}
	In the first benchmark test, 
	we investigate a simple elastic wave propagation in an elastic cable where an analytical solution is available for quantitative comparison. 
	This problem is also studied by Refs. \cite{dyka1995approach,vidal2007stabilized,bonet2001remarks,lee2016new} by using either 
	updated or total Lagrangian SPH method. 
	Following Refs. \cite{dyka1995approach,vidal2007stabilized,bonet2001remarks,lee2016new}, 
	we consider a rod of dimensions $10 \times 0.2 \times 0.2~ (m)$ with the left end fixed and the right end free as shown in Figure \ref{figs:cable-setup}. 
	The wave propagation is initialized by  imposing a velocity $v = 5 ~\text{m} / \text{s}$ along the length direction on the right quarter of the rod.
	We consider a linear elastic material of density $\rho = 8000 ~\text{kg} / \text{m}^3$, Young’s modulus $E = 200~ \text{GPa}$ and Poisson’s ratio $\nu = 0.0$. 
	
\begin{figure}[tb!]
		\centering
		\includegraphics[trim = 1mm 1mm 1mm 1mm, clip, width=\textwidth]{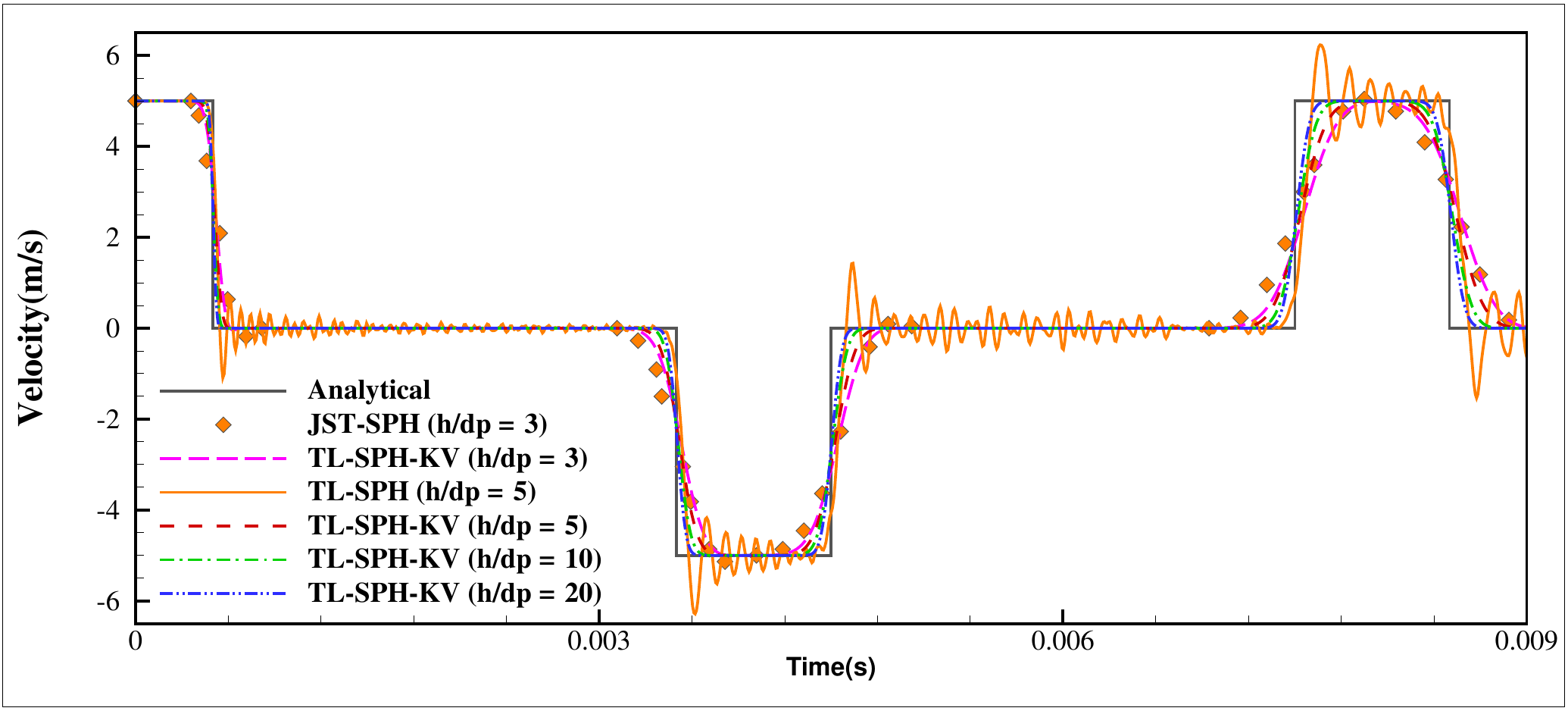} \\
		\includegraphics[trim = 1mm 1mm 1mm 2mm,clip, width=\textwidth]{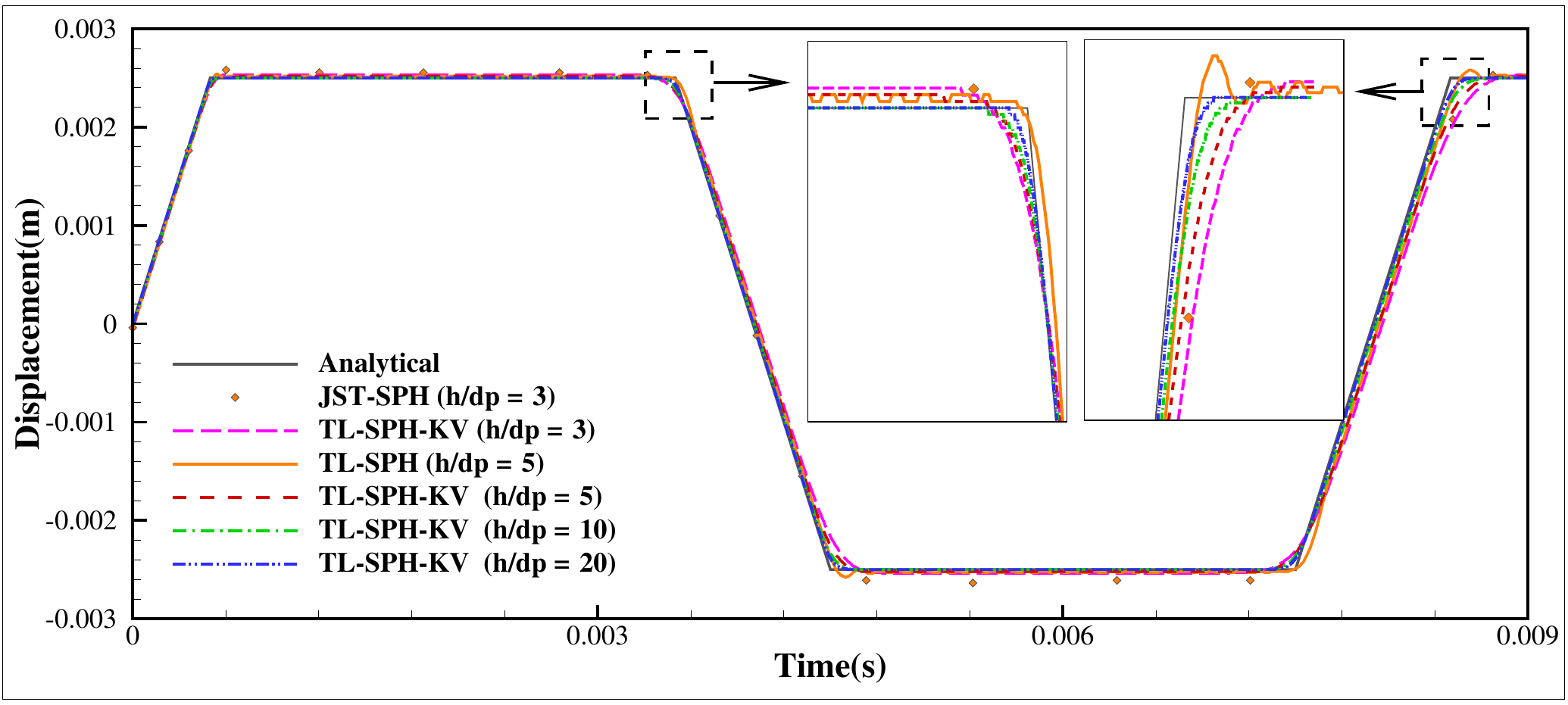} 
		\caption{Wave propagation in a cable: Velocity (upper panel) and displacement (bottom panel) profiles at the right tip end.
			A linear elastic material with density $\rho = 8000~ \text{kg} / \text{m}^3$, Young’s modulus $E = 200~ \text{GPa}$ and Poisson’s ratio $\nu = 0.0$ is applied.}
		\label{figs:cable-data}
	\end{figure}

	Figure \ref{figs:cable-data} shows the time histories of  velocity and displacement  
	in the length direction at the right tip end of the rod and the comparison with analytical solution. 
	Similar to the reports in Ref. \cite{bonet2001remarks}, 
	TL-SPH exhibits excessive oscillation,
	and similar overshoots in the velocity and displacement profiles, respectively.
	As expected, TL-SPH-KV predicts the correct velocity and displacement in both profiles. 
	The convergence study of the present method is also presented in Fig. \ref{figs:cable-data} 
	and shows clear convergence as the spatial resolution increases. 
	Note that, the comparison between the present results and those of Ref. \cite{lee2016new} is also reported in Fig.\ref{figs:cable-data}. 
	The main velocity and displacement plateaus of these results are in good agreement except that small overshoots are observed in those of Ref. \cite{lee2016new}. 
	%
	\subsection{Bending column}\label{subsec:bending-column}
	\begin{figure}[htb!]
		\centering
		\includegraphics[trim = 6cm 8cm 6cm 0.5cm, clip, width=.3\textwidth]{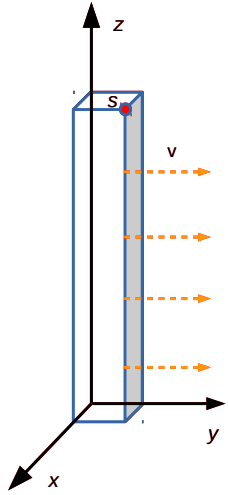}
		\caption{Bending column: Initial configuration. Note that point $S$ is located at $(1, 1, 6)^T\text{m}$.}
		\label{figs:bending-setup}
	\end{figure}
	In this second benchmark test, 
	we consider the bending deformation of a column of span $\text{L} = 6\text{m}$ and square cross section (height $h = 1 \text{m}$), 
	whose bottom end is clamped to the ground 
	and its body is allowed to bend freely 
	by imposing an initial uniform velocity $\mathbf{v} = (5\sqrt{3}, 5, 0)^T \text{m} \cdot \text{s}^{-1}$ as shown in Fig. \ref{figs:bending-setup}. 
	The neo-Hookean material with density $\rho = 1.1 \times 10^3 ~\text{kg} \cdot \text{m}^{-3}$, 
	Young's modulus $E=1.7\times10^7 ~\text{Pa}$ and 
	Poisson's ratio $\nu = 0.45$ is applied. 
	This test is also investigated by Aguirre et al.\cite{aguirre2014vertex} and we set their results as a reference for a rigorous comparison.
	
\begin{figure}[tb!]
		\centering
		\includegraphics[trim = 2cm 1mm 3cm 1mm, clip, width=0.8\textwidth]{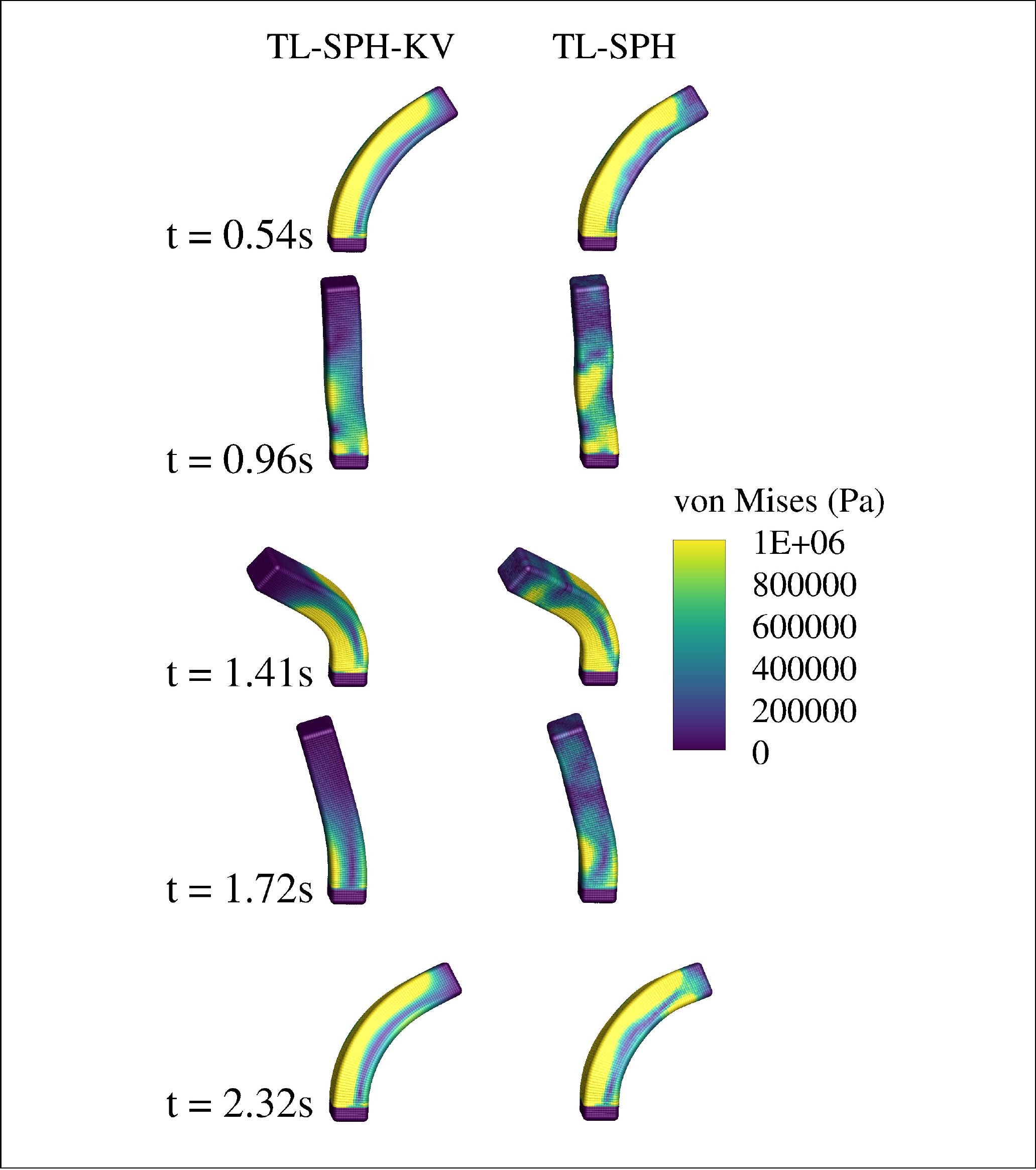}
		\caption{Bending column: Deformed configuration at different time stances for TL-SPH-KV and TL-SPH.
			The neo-Hookean material with density $\rho = 1.1 \times 10^3 ~\text{kg} \cdot \text{m}^{-3}$, Young's modulus $E=1.7\times10^7 ~\text{Pa}$ and 
			Poisson's ratio $\nu = 0.45$ is applied.}
		\label{figs:bending-particle}
	\end{figure}

\begin{figure}[tb!]
		\centering
		\includegraphics[trim = 1mm 1mm 1mm 1mm, clip, width=\textwidth]{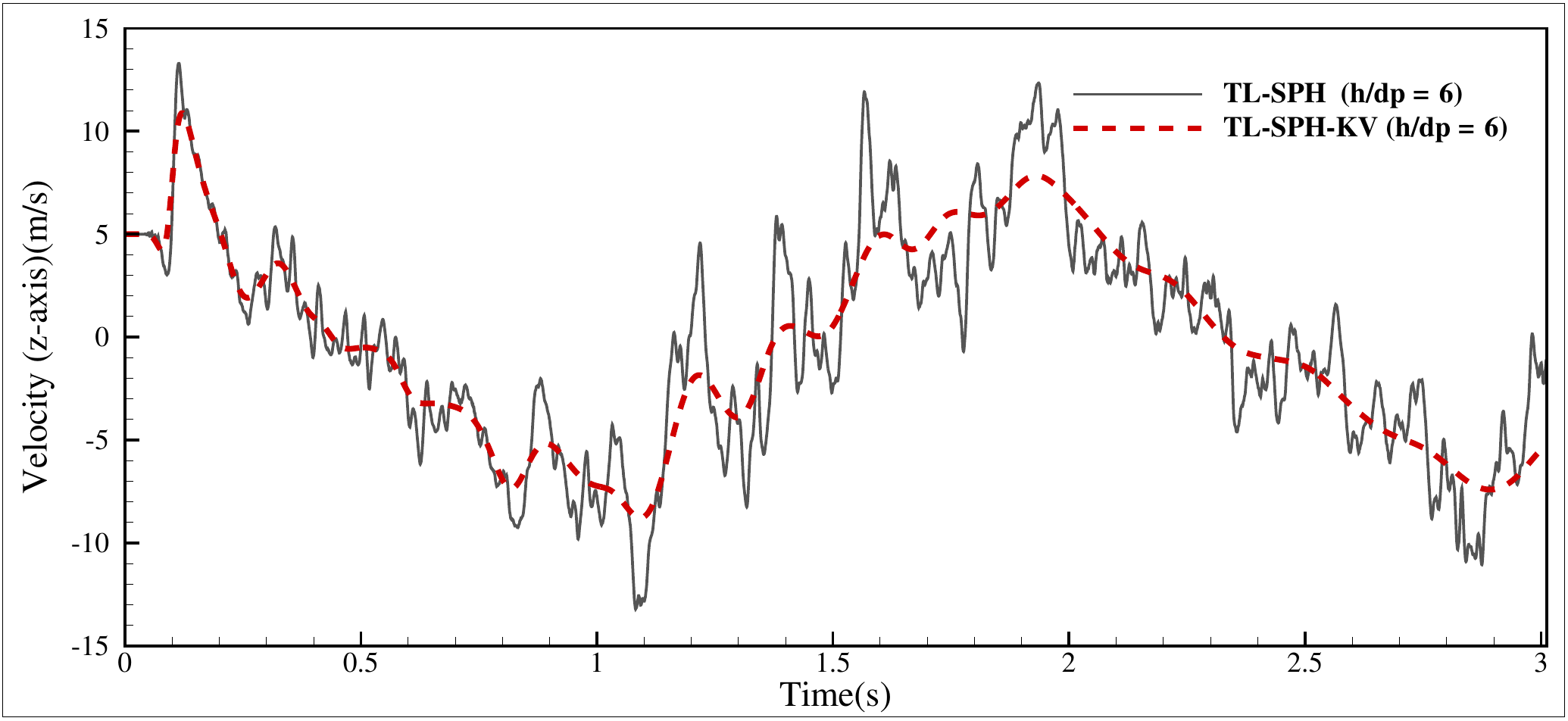} \\
		\includegraphics[trim = 1mm 1mm 1mm 1mm,clip, width=\textwidth]{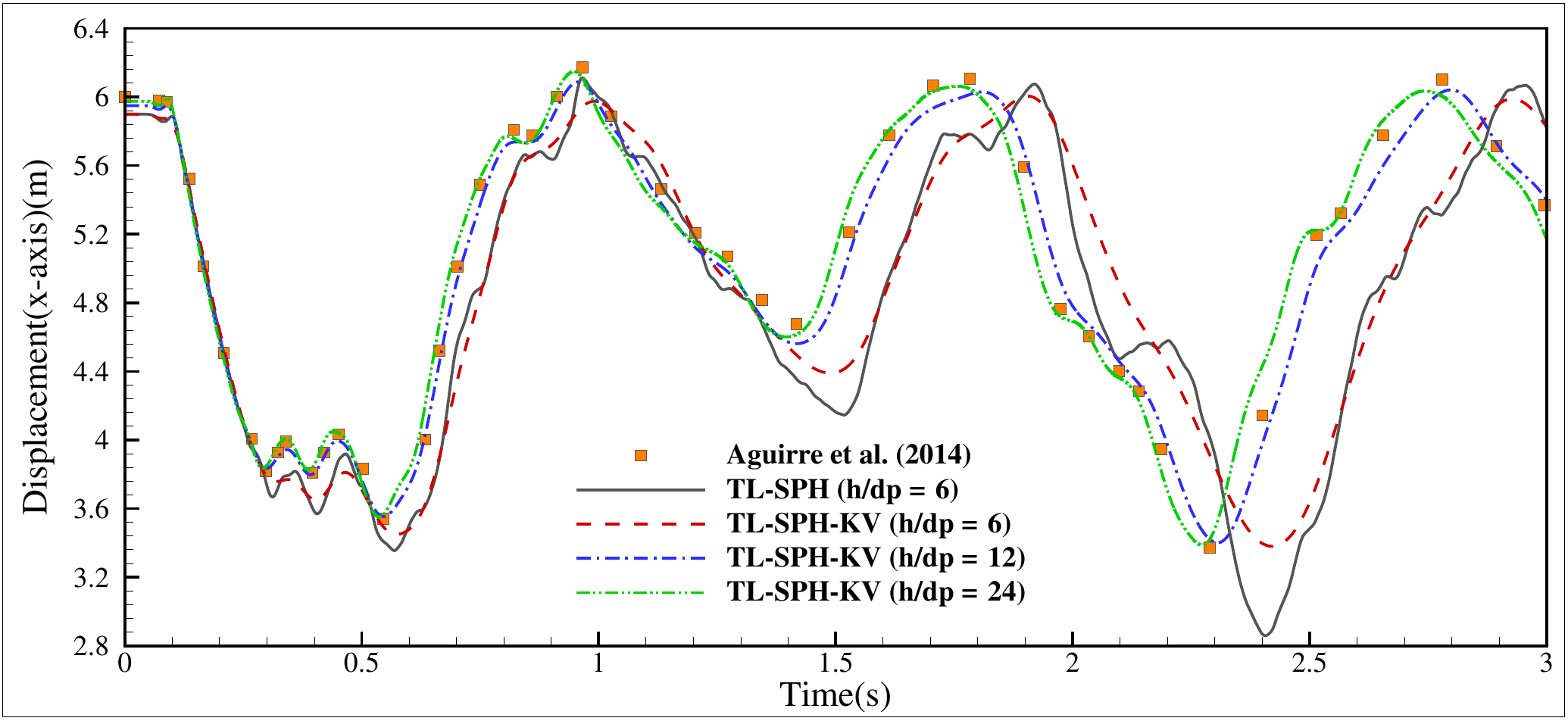} 
		\caption{Bending column: Velocity (upper panel) and displacement (bottom panel) profiles at the right tip end.
			The displacement profile is compared with data in Ref. \cite{aguirre2014vertex} and convergence study is also presented.
			The neo-Hookean material with density $\rho = 1.1 \times 10^3 ~\text{kg} \cdot \text{m}^{-3}$, Young's modulus $E=1.7\times10^7 ~\text{Pa}$ and 
			Poisson's ratio $\nu = 0.45$ is applied.}
		\label{figs:bending-data}
	\end{figure}

	Figure \ref{figs:bending-particle} shows the deformed configuration colored with von Mises stress contours for TL-SPH and TL-SPH-KV. 
	Obviously, oscillations in the von Mieses stress field is obtained by TL-SPH due to insufficient stabilization while these oscillations are suppressed by TL-SPH-KV.
	Similar to the last test, 
	TL-SPH produces noisy oscillations in the velocity profile which are eliminated by the present method as shown in Fig. \ref{figs:bending-data}. 
	Note that, Fig. \ref{figs:bending-data} also gives the time history of the vertical displacement of point $S$ and its comparison with that of Ref. \cite{aguirre2014vertex}. 
	Compared with the results reported in latter, 
	good agreements in the deformation are observed. 
	Also note that a 2nd-order convergence of the solution 
	is achieved by the present method with increasing spatial resolution,
	even though no linear reconstruction procedure is applied as in Ref. \cite{lee2019total}. 
	Compared with the present method, 
	the TL-SPH shows overshoots in the displacement profile which is 
	similar to fluctuations produced in the velocity field. 
	%
	\subsection{Twisting column}\label{subsec:twisting-column}
	\begin{figure}[htb!]
		\centering
		\includegraphics[trim = 5cm 1cm 5cm 0mm, clip, width=\textwidth]{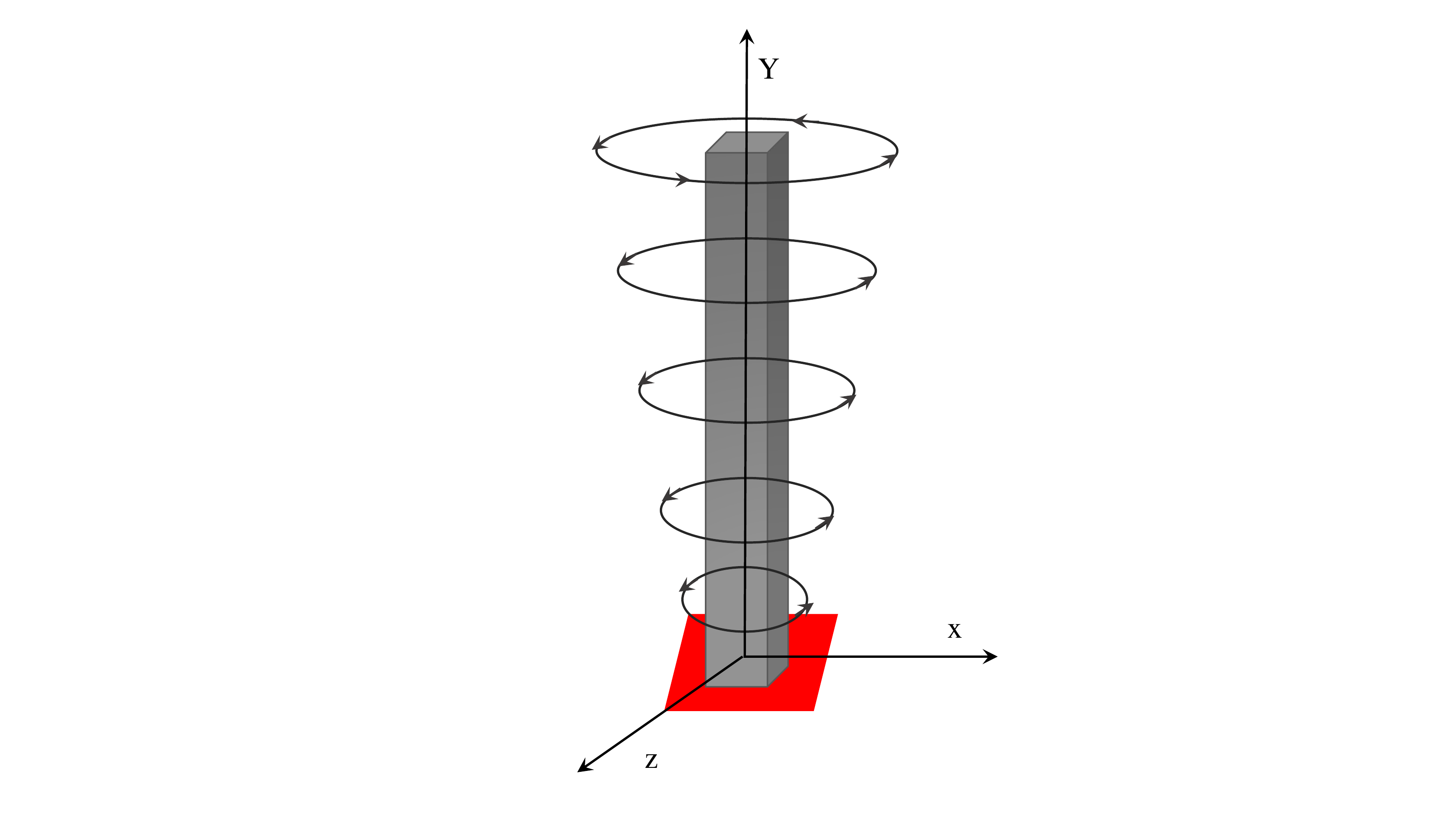}
		\caption{Twisting column: Initial configuration.}
		\label{figs:twisting-setup}
	\end{figure}
	Following Refs.\cite{lee2016new,lee2019total,gil2014stabilised}, 
	the example of bending column reported in Section \ref{subsec:bending-column} can be extended to 
	a more challenging test to assess the robustness of TL-SPH-KV for predicting the extremely highly nonlinear deformations.  
	As shown in Fig. \ref{figs:twisting-setup},  
	the twisting column is also clamped on its bottom face and the body is initialized with a sinusoidal rotational velocity field relative to the origin given by
	\begin{equation}
		\mathbf{v} \left( \mathbf{r} ^0 \right)  = \omega \times \mathbf{r} ^0, \omega = \left[  0.0, \Omega_0 \sin\left( \frac{\pi y}{2L}\right) , 0.0 \right]^T (\text{rad} \cdot \text s^{-1}) .
	\end{equation}
	The column material is modeled as nearly incompressible  by using a neo-Hookean constitutive model with density $\rho_0 = 1100 ~\text{Kg} / \text{m}^3$, 
	Young’s modulus $E =0.017 ~\text{GPa}$ and Poisson’s ratio $\nu = 0.4995$.
	
\begin{figure}[htb!]
		\centering
		\includegraphics[trim = 4cm 2cm 4cm 5mm, clip, width=0.85\textwidth]{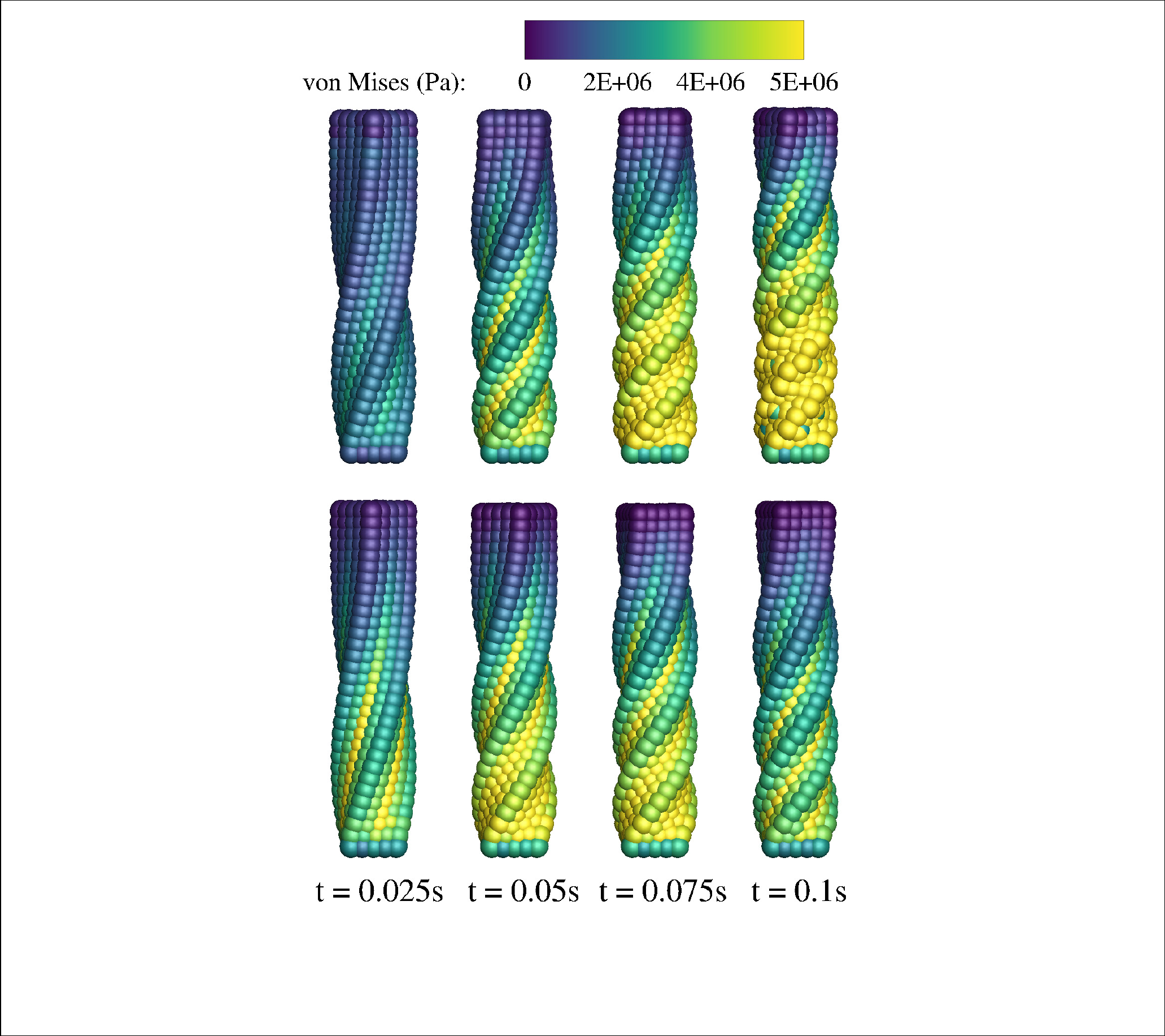} 
		\caption{Twisting column: Comparison of deformed configuration plotted with von Mises stress at serial time instance using the TL-SPH (upper panel) and 
			the TL-SPH-KV (bottom panel).
			Results obtained with a initial sinusoidal rotational velocity $\Omega_ = 105 ~\text{rad} / \text{s}$. 
			A neo-Hookean material with density $\rho_0 = 1100 ~\text{Kg} / \text{m}^3$, 
			Young’s modulus $E =0.017 ~\text{GPa}$ and Poisson’s ratio $\nu = 0.4995$ is applied.}
		\label{figs:twisting-particle}
	\end{figure}

\begin{figure}[htb!]
		\centering
		\includegraphics[trim = 1mm 1mm 1mm 1mm, clip, width=0.9\textwidth]{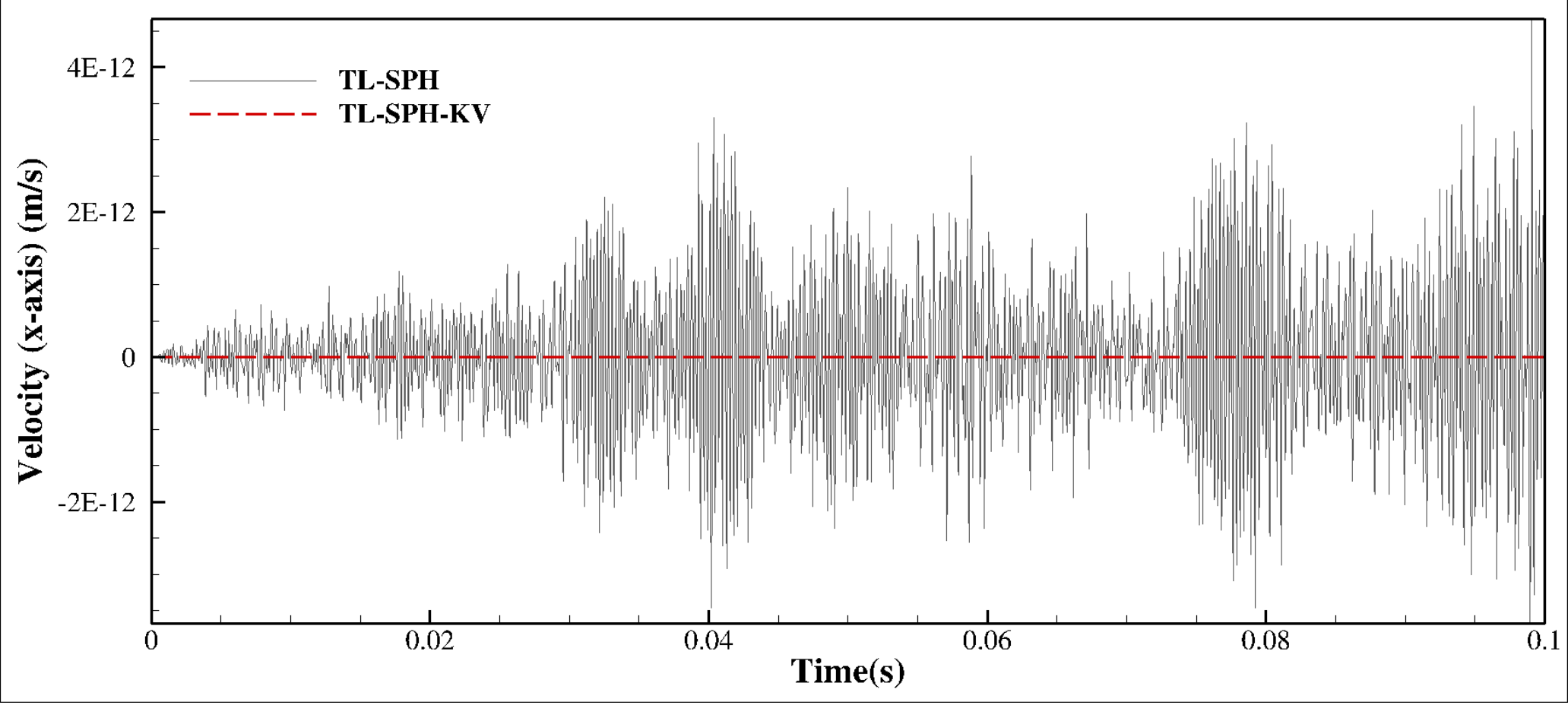}
		\includegraphics[trim = 1mm 1mm 1mm 1mm, clip, width=0.9\textwidth]{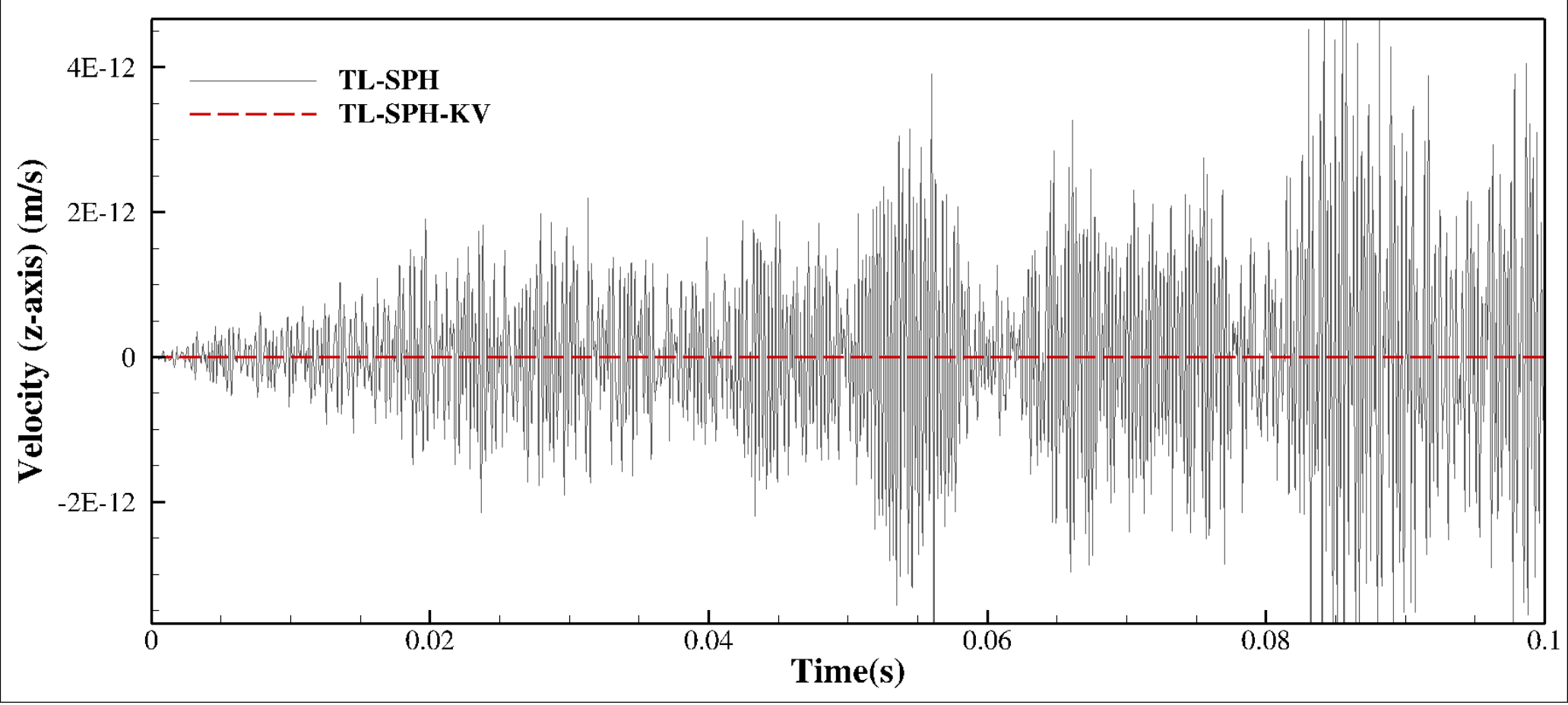}
		\caption{Twisting column: Time history of the velocity at the middle point of the free end using TL-SPH and TL-SPH-KV. 
			Results obtained with a initial sinusoidal rotational velocity $\Omega = 105 ~\text{rad} / \text{s}$. 
			A neo-Hookean material with density $\rho_0 = 1100~ \text{Kg} / \text{m}^3$, 
			Young’s modulus $E =0.017 ~\text{GPa}$ and Poisson’s ratio $\nu = 0.4995$ is applied.}
		\label{figs:twisting-velocity}
	\end{figure}

\begin{figure}[htb!]
		\centering
		\includegraphics[trim = 4cm 4cm 4cm 1cm, clip, width=0.85\textwidth]{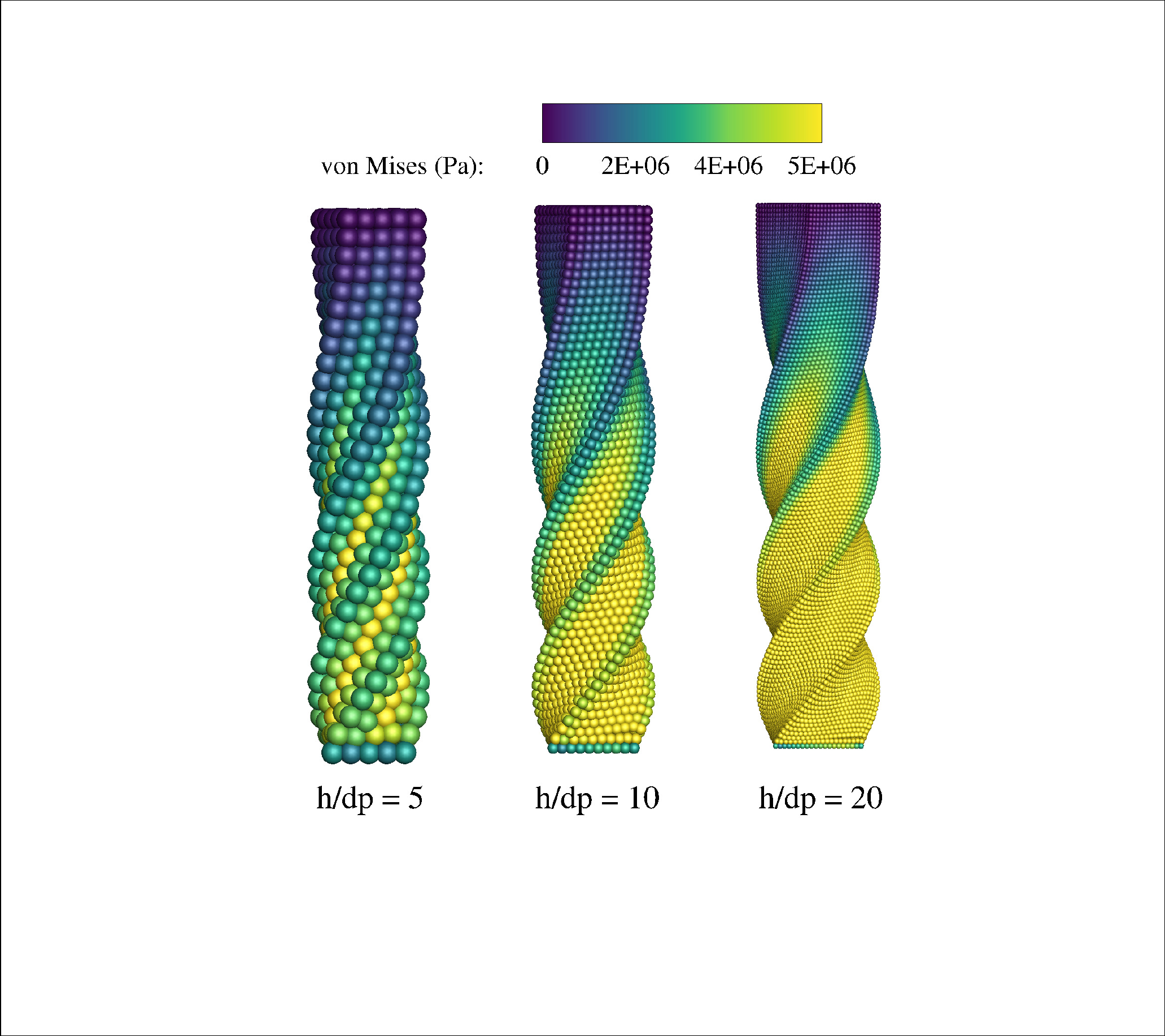}
		\caption{Twisting column: A sequence of particle refinement analysis using TL-SPH-KV.
			Results obtained with a initial sinusoidal rotational velocity $\Omega =105~ \text{rad} / \text{s}$. 
			A neo-Hookean material with density $\rho_0 = 1100 ~\text{Kg} / \text{m}^3$, 
			Young’s modulus $E =0.017 ~\text{GPa}$ and Poisson’s ratio $\nu = 0.4995$ is applied.}
		\label{figs:twisting-convergence}
	\end{figure}

	Figure \ref{figs:twisting-particle} shows the deformed configurations colored by von Mises stress with $\Omega_0 = 105 \text{rad} \cdot \text s^{-1}$
	obtained by TL-SPH and TL-SPH-KV. 
	Clearly, 
	TL-SPH produces non-physical stress fluctuations due to insufficient numerical stabilization and 
	then fails to capture the correct deformation pattern. 
	On the contrary, 
	TL-SPH-KV alleviates these discrepancies and predicts more accurate deformation patterns as those 
	in the literature \cite{lee2016new,lee2019total,gil2014stabilised} (see Figs 23 in Ref. \cite{lee2019total}), 
	demonstrating the robustness of the proposed method.
	Note that the results of  Ref. \cite{lee2019total} (see their Fig. 23) 
	reports about slightly more twist 
	may be due to the slight different neo-Hookean material models. 
	It is worth noting that TL-SPH-KV preserves the axial rotation very well without introducing out-of-axis characteristics, 
	as shown in Fig. \ref{figs:twisting-velocity} monitoring the time history of the horizontal velocity components 
	at the point $\mathbf{r} = \left[ 0.0, 6.0, 0.0\right]^T $. 
	As observed, TL-SPH generates much larger out-of-axis fluctuations. 
	Figure \ref{figs:twisting-convergence} shows the convergence study with particle refinement. 
	Both the deformation and von Mises stress resolution obtained 
	exhibit good convergence property for TL-SPH-KV. 
	
\begin{figure}[htb!]
		\centering
		\includegraphics[trim = 3cm 5.5cm 3cm 6mm, clip, width=0.95\textwidth]{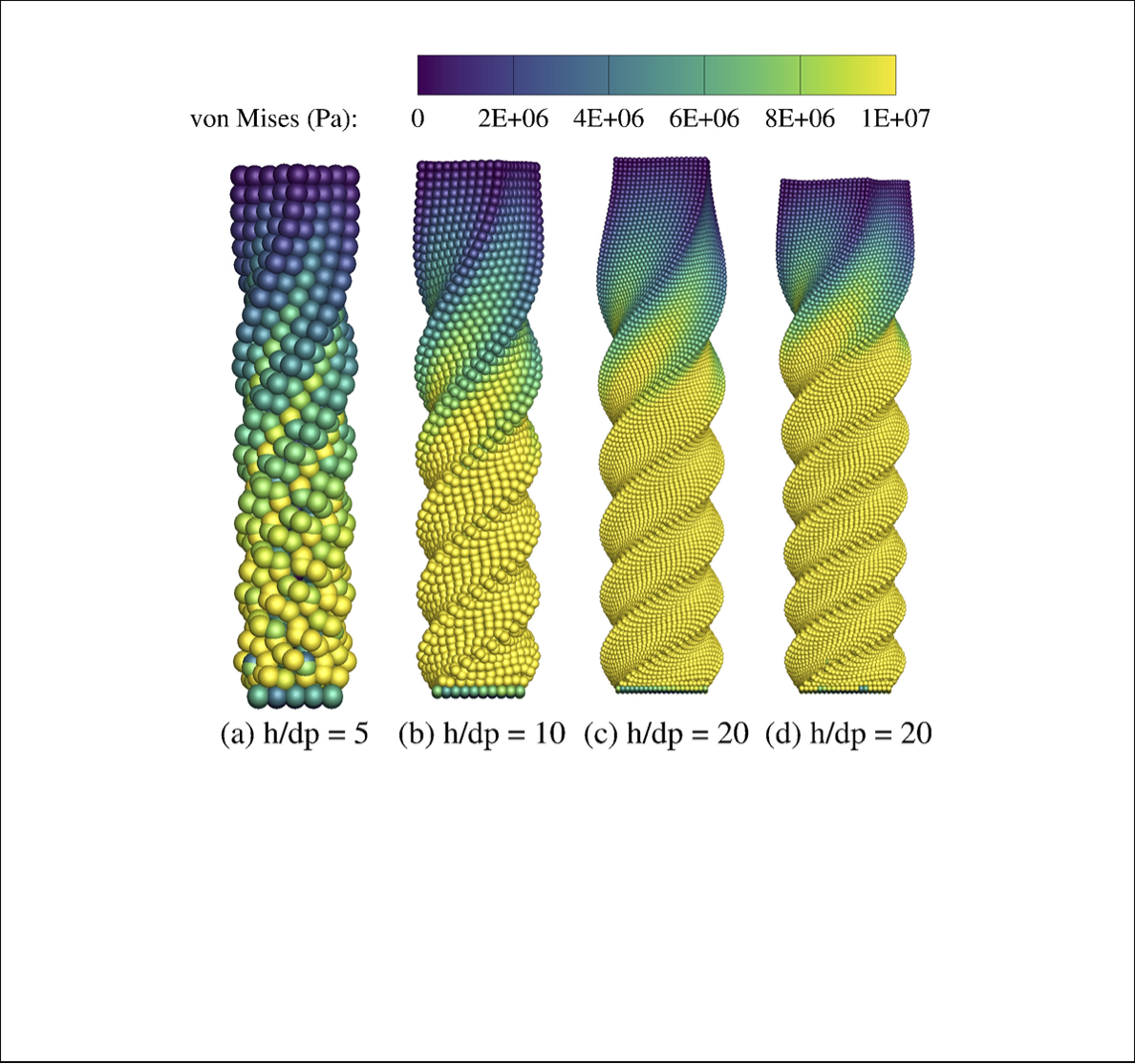} 
		\caption{Twisting column: Deformed configuration plotted with von Mises stress for large initial rotational velocity 
			obtained by using the  present TL-SPH-KV. 
			(a) - (c) particle refinement analysis for $\Omega_0 = 200 ~\text{rad} / \text{s}$ and (d) for $\Omega_0 = 300 ~\text{rad} / \text{s}$. 
			A neo-Hookean material with density $\rho_0 = 1100 ~\text{Kg} / \text{m}^3$, 
			Young’s modulus $E =0.017 ~\text{GPa}$ is applied. 
			Note that Poisson’s ratio is set as $\nu = 0.4995$ for (a) - (c) and $\nu = 0.49995$ for (d).}
		\label{figs:twisting-particle-2}
	\end{figure}

	To demonstrate the robustness of the present TL-SPH-KV, 
	we consider more challenging tests by increasing 
	the initial rotational velocity to $\Omega_0 = 200 ~\text{rad} / \text{s}$ and 
	$\Omega_0 = 300 ~\text{rad} / \text{s}$ (with Poisson's ratio $\nu = 0.49995$)  which induces extreme deformation and 
	leads to high requirement for robustness. 
	It is observed from Fig. \ref{figs:twisting-particle-2} (a) - (c) 
	that the extremely large deformation is well captured and, 
	numerical convergence is achieved with the increase of spatial resolution. 
	Also, with the even higher initial angular velocity, 
	one more twist is obtained as shown in Fig. \ref{figs:twisting-particle-2} (d), 
	which gives high-resolution results on the deformed configuration. 
	%
	\subsection{Stent structures}\label{subsec:stent}
	\begin{figure*}
		\centering
		\begin{subfigure}[b]{0.48\textwidth}
			\includegraphics[trim = 2mm 2mm 2mm 2mm, clip,width=.95\textwidth]{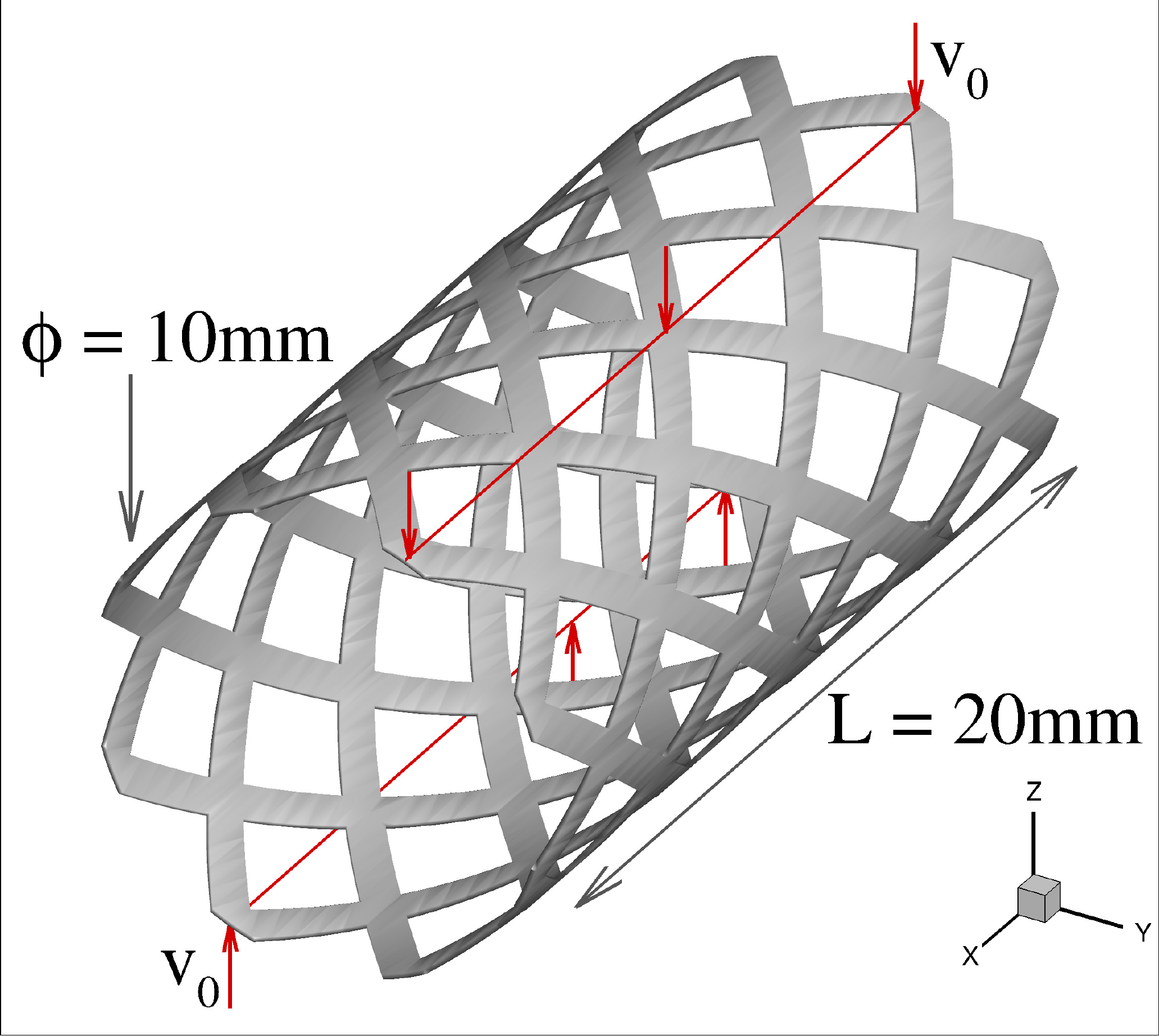}
			\caption{PS-shaped stent}
			\label{fig:ps-stent}
		\end{subfigure}
		\begin{subfigure}[b]{0.48\textwidth}
			\vspace{0.5cm}
			\includegraphics[trim = 2mm 2mm 2mm 2mm, clip,width=.95\textwidth]{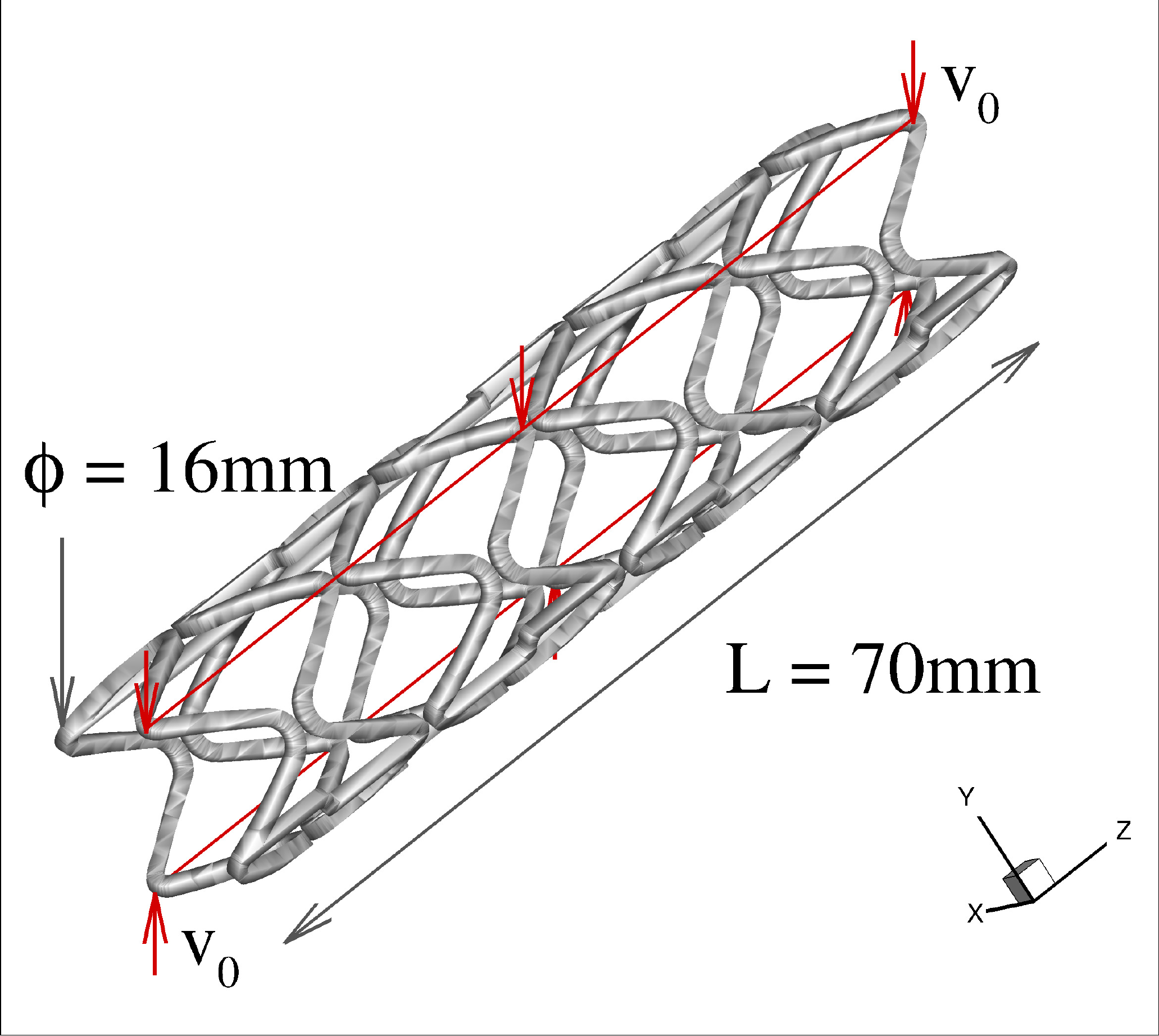}
			\caption{C-shaped stent}
			\label{fig:c-stent}
		\end{subfigure}
		\caption{Computer-aided Design (CAD) geometries of stent structures: 
			Palmaz-Schatz shaped (PS-shaped) stent and Cypher shaped (C-shaped) stent. 
			The corresponding CAD files in STL format can be downloaded from our code repository or GrabCAD.}
		\label{figs:stent-setup}
	\end{figure*}
	In the last example, 
	complex flexible structures are investigated to demonstrate the robustness and versatility of TL-SPH-KV. 
	As shown in Fig. \ref{figs:stent-setup}, two different stent structures, viz. 
	Palmaz-Schatz shaped (PS-shaped) and Cypher shaped (C-shaped), are considered. 
	Note that the present stent structures are realistic cardiovascular stent and widely used in biomedical applications.
	To study the deformation pattern of the stent structures,
	we apply a velocity field with magnitude of $v_0 = 5.0 \text m \cdot \text s^{-1}$ at the top and bottom of the structure, 
	which is modeled as neo-Hookean material with density $\rho_0 = 1100 ~\text{Kg} / \text{m}^3$,
	Young’s modulus $E =0.017 ~\text{GPa}$ and Poisson’s ratio $\nu = 0.45$ . 
	
\begin{figure}[htb!]
		\centering
		\includegraphics[trim = 1mm 1mm 1mm 1mm, clip,width=\textwidth]{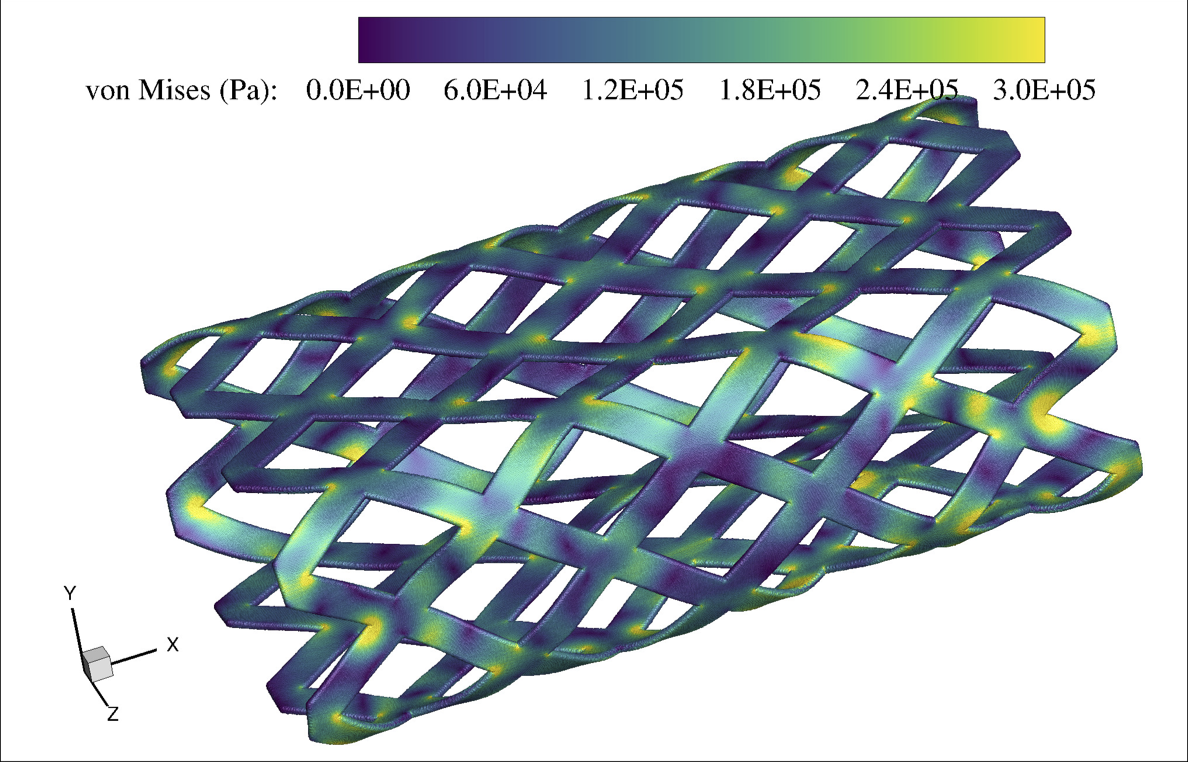}\\
		\includegraphics[trim = 1mm 4cm 1mm 1mm, clip,width=0.485\textwidth]{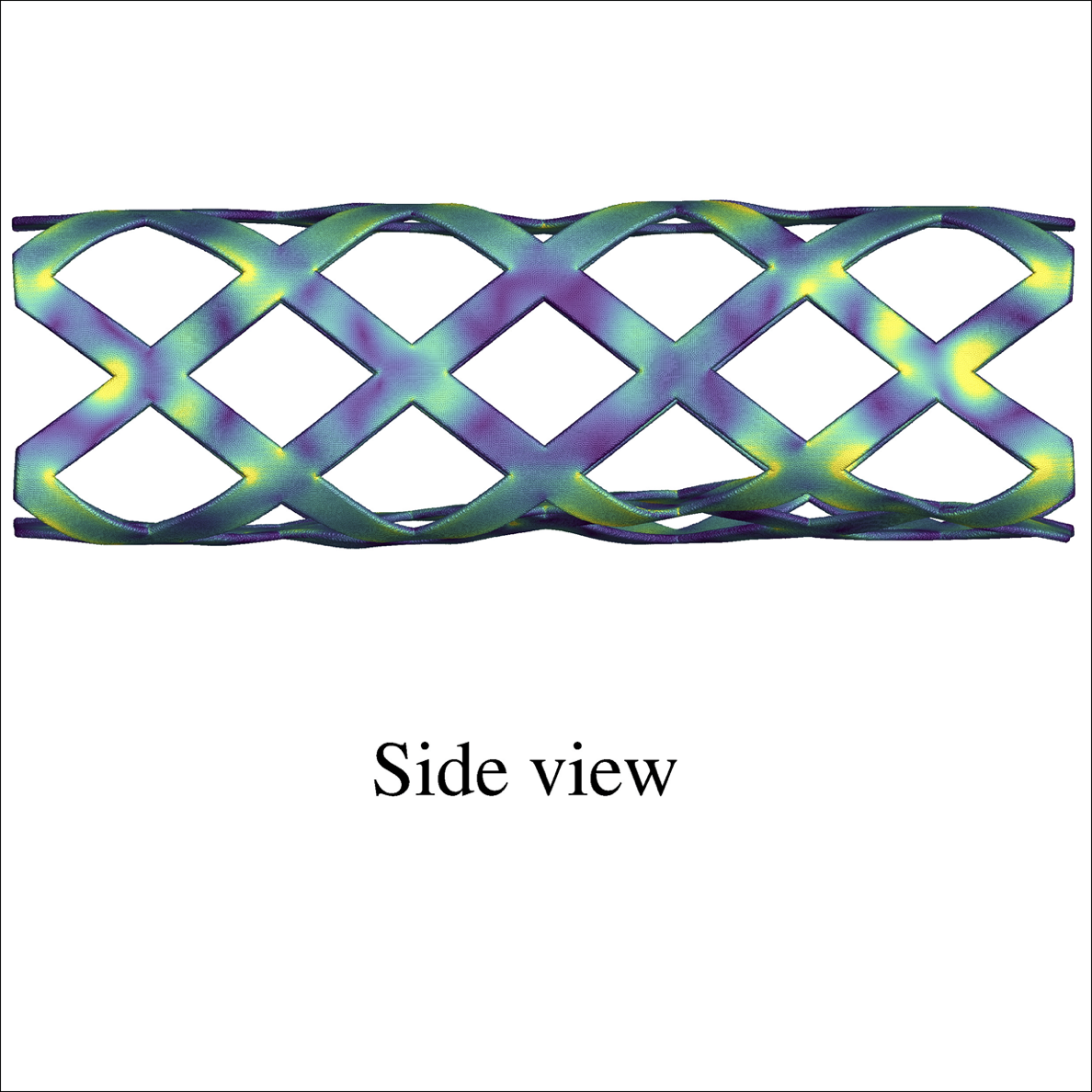}
		\includegraphics[trim = 1mm 4cm 1mm 1mm, clip,width=0.485\textwidth]{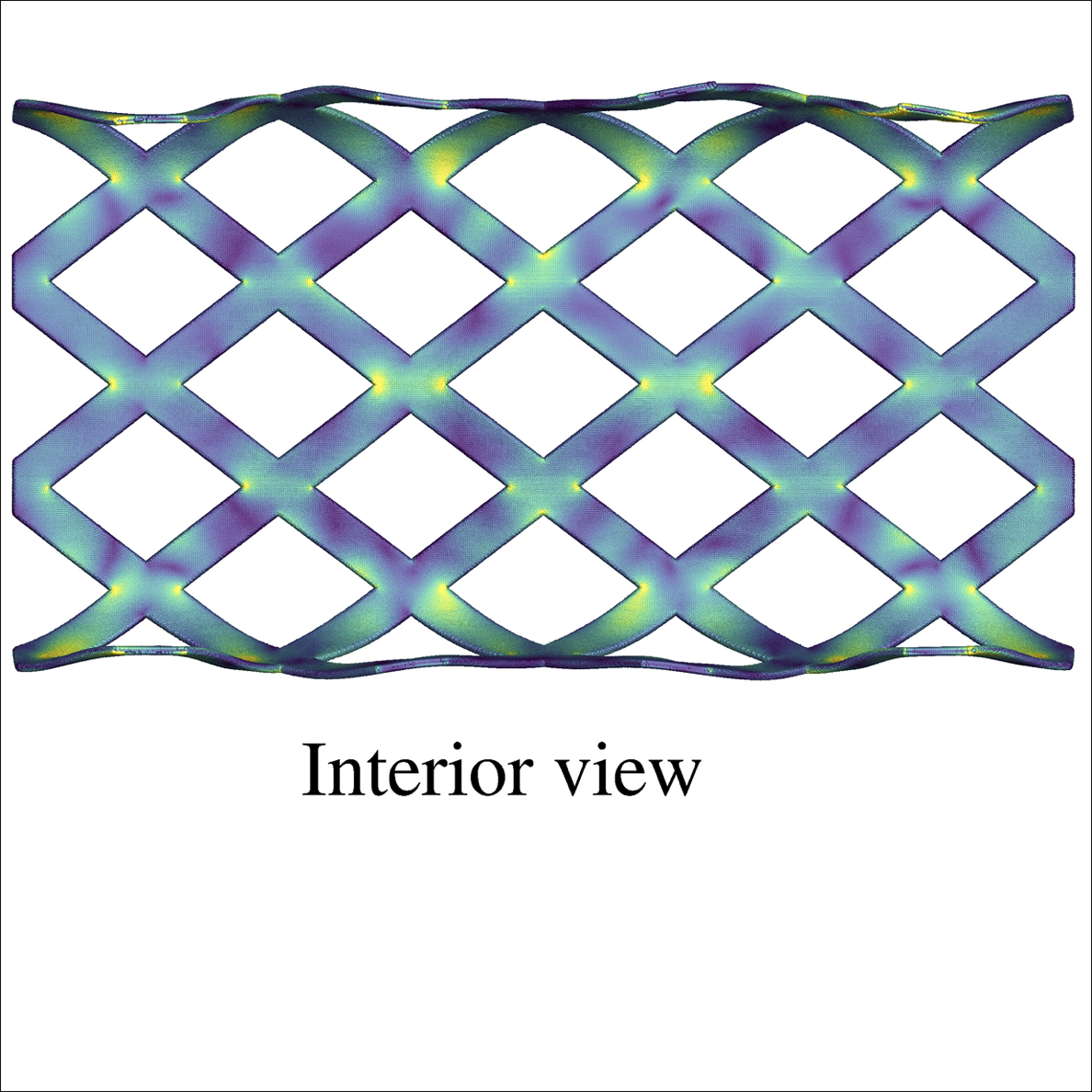}
		\caption{PS-shaped stent structure: Deformed configuration with von Mises stress contour at $t = 0.02 ~s$ with present TL-SPH-KV.
			Here, $\mathbf{v}_0 = 5.0 ~\text{m} / \text{s}$ is applied to impose the initial condition and the neo-Hookean material with 
			density $\rho_0 = 1100 ~\text{Kg} / \text{m}^3$, Young’s modulus $E =0.017 ~\text{GPa}$ and Poisson’s ratio $\nu = 0.45$ is used.}
		\label{figs:ps-stent-deformed}
	\end{figure}

\begin{figure}[htb!]
		\centering
		\includegraphics[trim = 1mm 1mm 1mm 1mm, clip,width=\textwidth]{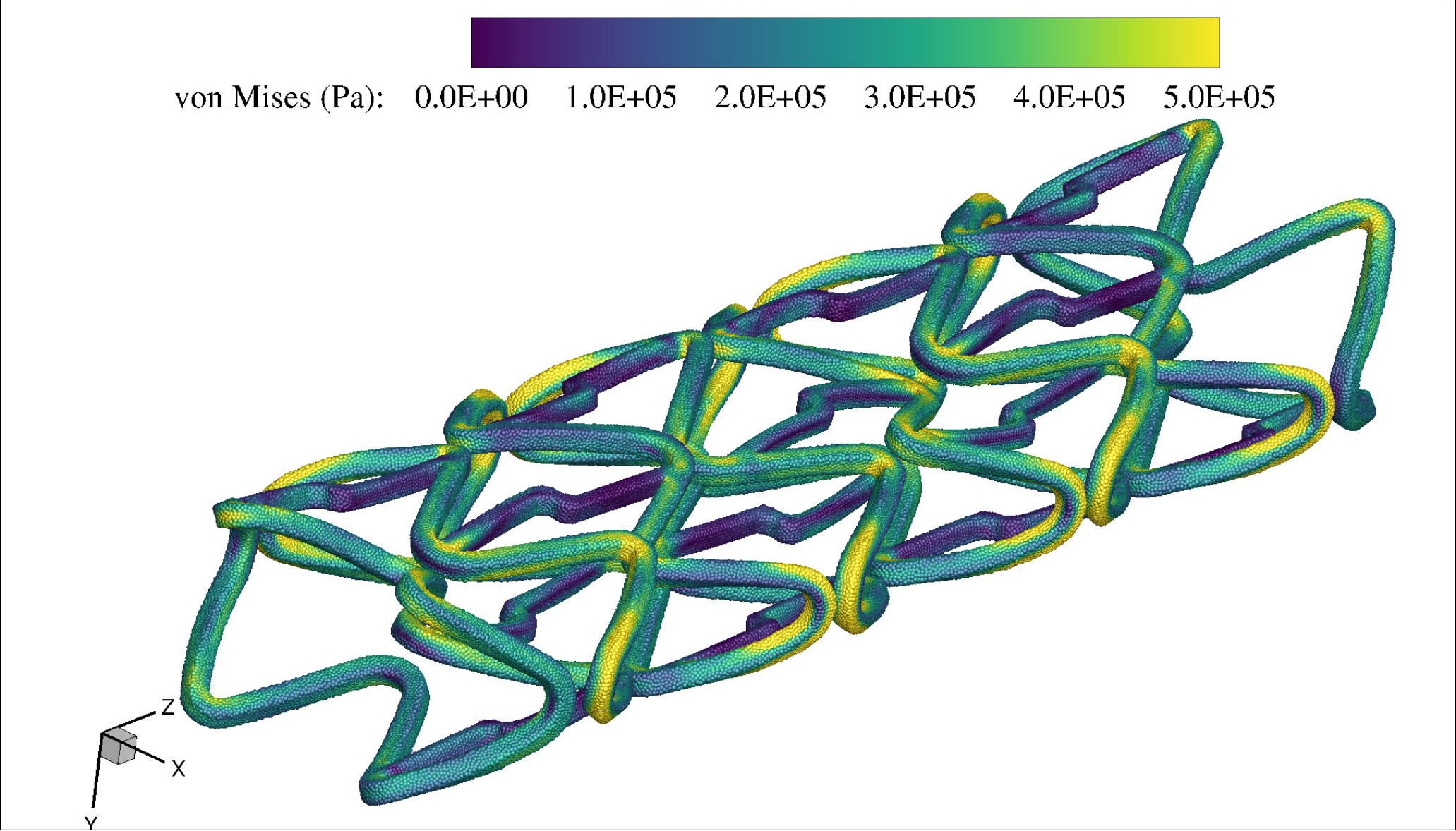}\\
		\includegraphics[trim = 1mm 4cm 1mm 1mm, clip,width=0.485\textwidth]{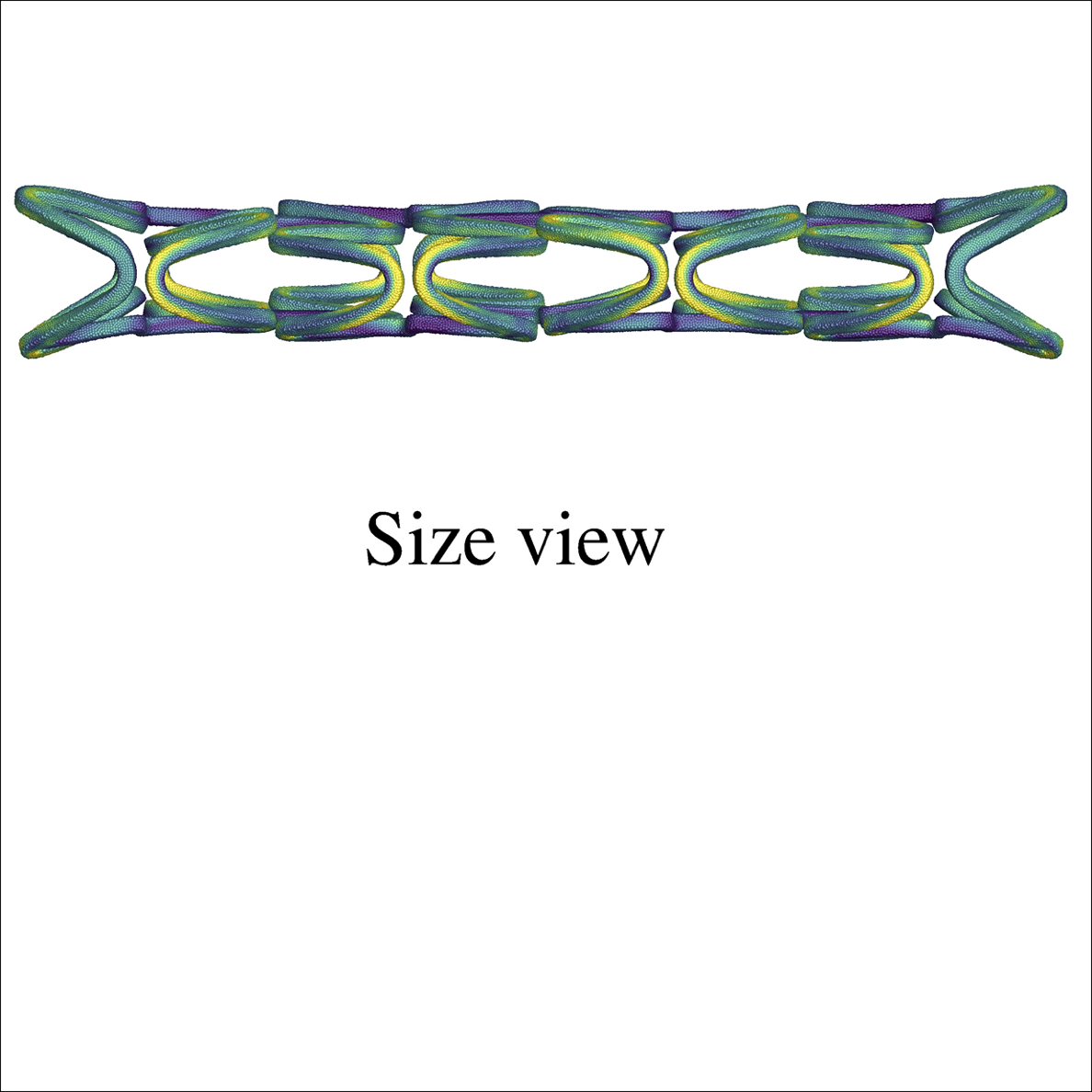}
		\includegraphics[trim = 1mm 4cm 1mm 1mm, clip,width=0.485\textwidth]{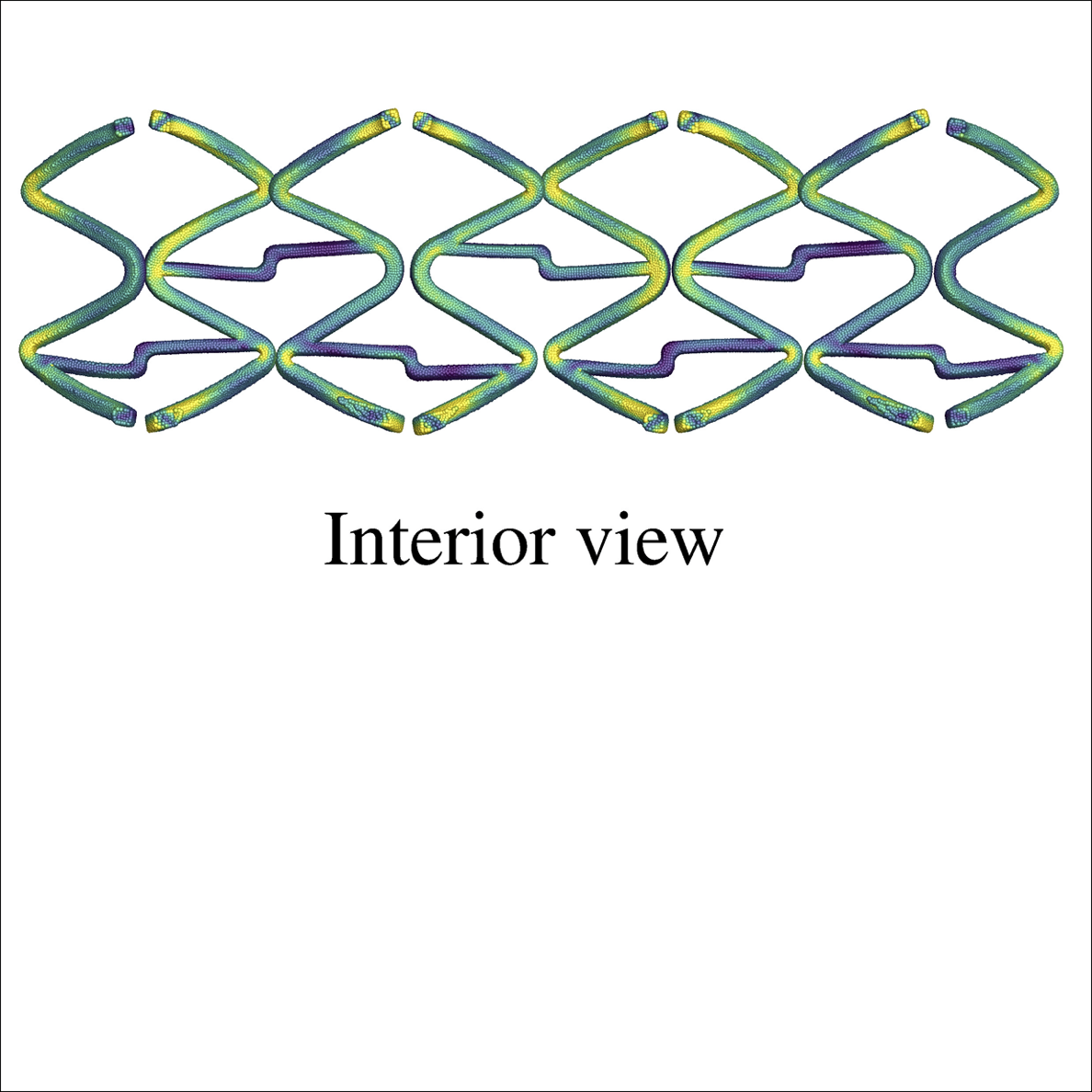}
		\caption{C-shaped stent structure: Deformed configuration with von Mises stress contour at $t = 0.025 ~s$ with present TL-SPH-KV.
			Here, $\mathbf{v}_0 = 5.0~ \text{m} / \text{s}$ is applied to impose the initial condition and the neo-Hookean material with 
			density $\rho_0 = 1100 ~\text{Kg} / \text{m}^3$, Young’s modulus $E =0.017 ~\text{GPa}$ and Poisson’s ratio $\nu = 0.45$ is used.}
		\label{figs:c-stent-deformed}
	\end{figure}

	Figure \ref{figs:ps-stent-deformed} 
	shows the overall deformation of the PS-shaped stent structure at time $t = 0.02 \text{s}$ with von Mises stress contour. 
	It can be observed that the regions of high stress are concentrated at the four corners of the cells 
	rather than in the middle of the struts or the bridging strut itself. 
	This is due to the fact that the struts pull apart from each other to form a rhomboid shape of cells during the deformation. 
	Figure \ref{figs:c-stent-deformed} 
	presents the deformed configuration of the C-shaped stent structure at time $t = 0.025 \text{s}$ with von Mises stress contour. 
	Different with the PS-shaped stent, 
	the regions of high stress in C-shaped stent are concentrated at the curved regions of the struts. 
	Another notable difference is the obvious expansion of the free-end strut in the C-shaped stent because of the antisymmetric constraint link. 
	Generally, it is remarkable that for both structures the von Mises stress field is reasonably captured. 
	To the best knowledge of the authors, 
	this is first time that a SPH-based method is successfully extended to the simulation of realistic cardiovascular stent and this will open up interesting possibilities for modeling bio-mechanical applications, 
	where this consideration is very relevant.
	%
	%
	%
	\section{Concluding remarks}\label{sec:conclusion}
	In this paper,
	we present a simple and robust artificial damping method for stabilize the TL-SPH simulation of the solid mechanics problems involving large deformations. 
	The proposed stabilization strategy is based on a Kelvin-Voigt type damper 
	in the constitution equation and a scaling factor imitating 
	a von Neumann-Richtmyer type artificial viscosity. 
	A set of numerical examples together with very challenging cases 
	have been investigated, and it is shown that the present method is free of the non-physical 
	fluctuations suffered by the original TL-SPH. 
	Finally, its versatility is also demonstrated by the simulation of complex stent structures, which is a stepping stone to possible applications in the field of bio-mechanics. 
	%
	%
	\section*{CRediT authorship contribution statement}
	{\bfseries  Chi Zhang:} Investigation, Conceptualization, Methodology, Visualization, Validation, Formal analysis, Writing - original draft, Writing - review \& editing; 
	{\bfseries  Yujie Zhu:} Investigation, Writing - review \& editing;
	{\bfseries  Yongchuan Yu:} Investigation; 
	{\bfseries  Massoud Rezavand:} Investigation, Writing - review \& editing; 
	{\bfseries  Xiangyu Hu:} Supervision, Conceptualization, Methodology, Investigation, Writing - review \& editing.
	%
	%
	\section*{Declaration of competing interest }
	The authors declare that they have no known competing financial interests 
	or personal relationships that could have appeared to influence the work reported in this paper.
	%
	%
	\section{Acknowledgement}
	The authors would like to express their gratitude to Deutsche Forschungsgemeinschaft (DFG) 
	for their sponsorship of this research under grant numbers 
	DFG HU1527/10-1 and HU1527/12-1.
	%
	%
	\clearpage
	\bibliography{mybibfile}
	%
	%
\end{document}